\documentclass[sigconf]{acmart}
\usepackage[most]{tcolorbox}
\AtBeginDocument{%
  \providecommand\BibTeX{{%
    \normalfont B\kern-0.5em{\scshape i\kern-0.25em b}\kern-0.8em\TeX}}}

\copyrightyear{2022}
\acmYear{2022}
\acmConference[ICPC '22]{30th International Conference on Program Comprehension}{May 16--17, 2022}{Virtual Event, USA}
\acmBooktitle{30th International Conference on Program Comprehension (ICPC '22), May 16--17, 2022, Virtual Event, USA}

\acmConference[ICPC 2022]{The 30th International Conference on Program Comprehension}{May 21–22, 2022}{Pittsburgh, PA, USA}
 \newcommand{\yc}[1]{\textcolor{black}{#1}}

\begin{document}

\title{On the Effectiveness of Pretrained Models for API Learning}


\author{Mohammad Abdul Hadi}
\affiliation{%
  \institution{University of British Columbia}
  \streetaddress{333 University Way}
  \city{Kelowna}
  \country{Canada}}
\email{mohammad.hadi@ubc.ca}

\author{Imam Nur Bani Yusuf}
\affiliation{%
  \institution{Singapore Management University}
  \city{}
  \country{Singapore}
}
\email{imamy.2020@phdcs.smu.edu.sg}

\author{Ferdian Thung}
\affiliation{%
  \institution{Singapore Management University}
  \city{}
  \country{Singapore}
}
\email{ferdianthung@smu.edu.sg}

\author{Kien Gia Luong}
\affiliation{%
  \institution{Singapore Management University}
  \city{}
  \country{Singapore}
}
\email{kiengialuong@smu.edu.sg}

\author{Jiang Lingxiao}
\affiliation{%
  \institution{Singapore Management University}
  \city{}
  \country{Singapore}
}
\email{lxjiang@smu.edu.sg}

\author{Fatemeh H. Fard}
\affiliation{%
  \institution{University of British Columbia}
  \streetaddress{333 University Way}
  \city{Kelowna}
  \country{Canada}}
  \email{fatemeh.fard@ubc.ca}

\author{David Lo}
\affiliation{%
  \institution{Singapore Management University}
  \city{}
  \country{Singapore}
}
\email{davidlo@smu.edu.sg}

\renewcommand{\shortauthors}{Hadi et al.}

\begin{abstract}
    Developers frequently use APIs to implement certain functionalities, such as parsing Excel Files, reading and writing text files line by line, etc. Developers can greatly benefit from automatic API usage sequence generation based on natural language queries for building applications in a faster and cleaner manner. Existing approaches utilize information retrieval models to search for matching API sequences given a query or use \yc{RNN-based} 
    encoder-decoder 
    to generate API sequences. As it stands, the first approach treats queries and API names as bags of words. It lacks deep comprehension of the semantics of the \yc{queries}. The latter approach adapts a neural language model to encode a user query into a fixed-length context vector and generate API sequences from the context vector.
  
    We want to understand the effectiveness of recent Pre-trained Transformer based Models (PTMs) for the API learning task. These PTMs are trained on large natural language corpora in an unsupervised manner to retain contextual knowledge about the language and have found success in solving similar Natural Language Processing (NLP) problems. However, the applicability of PTMs has not yet been explored for \yc{the} API sequence generation \yc{task}. We use a dataset that contains 7 million annotations collected from GitHub to evaluate the PTMs empirically. This dataset was also used to assess previous approaches. Based on our results, PTMs generate more accurate API sequences and outperform other related methods by $\sim$11\%. We have also identified two different tokenization approaches that can contribute to a significant boost in PTMs' performance for the API 
    \yc{sequence generation task}.
\end{abstract}





\maketitle

\section{Introduction}
Developers frequently rely on existing class libraries to implement particular 
\yc{functionalities} by invoking APIs that correspond to the libraries. It is extremely helpful to identify which APIs to use and in what order developers should invoke the API-related methods. For developers, who may need to face a steep learning curve, learning and becoming acquainted with an unfamiliar library can be challenging. It is not uncommon for a library as large as JDK to have hundreds or thousands of APIs, but they are not well documented regarding the usage patterns of API methods. Our literature study reveals that inadequate or absent resources cause obstacles for learning APIs, and a major challenge for API users is to discover the subset of the APIs that can help complete a task.

Search engines, such as Google and Bing, are commonplace to discover APIs and their usage sequences. However, the general web search engines are not designed to cater specifically to programming or code-related tasks, so they are often deemed ineffective when it comes to assisting with such tasks. Currently, there is no easy way for developers to learn about APIs and their usage sequences from the web pages returned by the search engines, which the search engines retrieved based on keyword matching without considering the semantics of natural language queries, making it more difficult to locate relevant code snippets and APIs associated with them.


Using a statistical word alignment model, Raghothaman et al. recently proposed SWIM, 
\yc{to translate} a natural language query into a list of possible APIs~\cite{swim}. To retrieve relevant API sequences, SWIM applies UP-Miner~\cite{tseng2015up}, a utility pattern mining toolbox, to a list of APIs. This method has a critical problem; 
it ignores the ordering of the natural language queries as these queries are treated as a bag of words. Therefore, it cannot recognize a deep semantic meaning within the natural language queries. This critical problem was addressed by 
DeepAPI~\cite{deep-api-learning}, a deep-learning-based method that translates a natural language query into relevant API usage sequences. API learning has been formulated as a machine translation problem: given a natural language query, the goal is to translate it into an API sequence. 
\yc{DeepAPI leverages RNN-based encoder-decoder architecture, and also adapts attention mechanism~cite(attention paper)}. \yc{It} has demonstrated a deep understanding of natural language queries in two aspects: learning the semantics of words by embedding them in context-vector representations of context and understanding the sequence of words in a natural language query and the associated APIs instead of focusing on word-to-word alignment.

DeepAPI 
\yc{is trained} on a corpus 
\yc{containing pairs of API sequences and the corresponding natural language queries.}
The encoder learns how to encode each \yc{query} into a fixed-length context vector for the decoder to use for decoding the API sequence. In the training phase, the 
\yc{DeepAPI} learns the semantics of 
\yc{the queries and the corresponding API sequences} 
and accordingly updates its parameters. 
\yc{After the training phase}, DeepAPI can generate the API sequences 
given a query describing the functionality of the intended API.

During the training phase 
, the encoder and the decoder 
\yc{in DeepAPI} is focused on learning: \yc{1)} the semantics of 
\yc{the queries in the input and the corresponding API sequences in the output, and 2)} the correlations between the input and the output sequences. We want to separate these two tasks and propose to leverage a Pre-trained Transformer model (PTM). \yc{PTM is a model that leverages Transformer encoder-decoder architecture(~cite transformer paper) that has been pre-trained using certain techniques on a data-rich corpora in the pre-training phase.} 
\yc{Because a PTM has been pre-trained on a data-rich corpora beforehand, it has a sense of understanding the semantic meaning of the queries and the corresponding API sequences. Intuitively, the learning burden of the model can be reduced if the model already has such a knowledge in the beginning of its training. Hence, the model can focus to learn the correlation between the input and the output sequences in the training phase.}

In our study, we aim to answer the following research questions:
\begin{itemize}
  \item[\textbf{RQ1:}] Can PTMs yield better performance than the prior approaches for API learning task?
  \item[\textbf{RQ2:}] Can different tokenization approaches help the PTMs perform better?
\end{itemize}

In the first research question, we explore how the current \yc{existing} PTMs perform compared to the other existing tools, such as SWIM and 
\yc{DeepAPI}. The existing tools are based on the curated code-search and pattern mining algorithms, and also deep learning-based approaches. 
\yc{Our results shows} that PTMs can outperform the prior approaches by $\sim$3\% to $\sim$7\% \yc{in terms of BLEU score}.

The second research question deals with whether we could boost the performance of PTMs by introducing different tokenization techniques. \yc{Tokenization is a technique to split a piece of text or sentence into a smaller unit called token <insert citation here>. Tokenization may affect the performance of a model because it affects how the input is represented to the model}. We have proposed two tokenization techniques. 
\yc{Leveraging our customized tokenization approach, our results demonstrate that} the PTMs \yc{can outperform the prior state of the art by} $\sim$11\% \yc{in terms of} BLEU Score.

The main contribution of our paper is stated below:
\begin{itemize}
  \item To the best of our knowledge, 
  \yc{we are the first who adapt PTMs for the API sequence generation task.} 
  \item \yc{We propose two novel} tokenization techniques to help improve the performance of the PTMs \yc{in the API sequence generation task.} 
  \item \yc{We evaluate the existing PTMs against existing approaches using a corpus containing 7 million pairs of Java API sequences and the corresponding queries used in the prior study <cite the deep api learning paper here>.} \yc{Our results demonstrate that PTMs can outperform the prior approaches by $\sim$3\% to $\sim$7\% in terms of BLEU score.Leveraging our tokenization technique, PTM can outperforms the prior best approach by $\sim$11\% \yc{in terms of} BLEU Score.}
  \item We provide a replication package and open source our codes and dataset to help fellow researchers and practitioners to reproduce our results. \footnote{https://github.com/Mohammad-Abdul-Hadi/PTM-API-Learning} 
\end{itemize}

Below is an outline of the remainder of this paper. Section \ref{sec:background_knowledge} introduces background knowledge. Sections \ref{sec:methodology} and \ref{sec:res} describe the methodology of our approach \yc{followed by the} results. Section \ref{sec:disc} 
discusses 
the interpretations and implications of our findings, along with the threats to the validity. Section \ref{sec:related_work} presents the related works. Finally, we conclude our study and present future work in Section \ref{sec:concluson}.

\section{Prelimineries}
\label{sec:background_knowledge}
In this section, we will discuss three different approaches used for API sequence generation from natural language queries: Information Retrieval Models, Deep Learning Approaches without Pretrained Knowledge, and Deep Learning Approaches with Pretrained Knowledge.

\subsection{Information Retrieval Models}

\subsubsection{Combination of Code Search and Pattern Mining for API Sequence generation}
Researchers have used code search using information retrieval techniques \cite{Intelligent-Code-Search-Engine, sourcerer, codehow, portfolio} to identify API sequences in a code corpus that correspond to a given query, followed by the use of an API usage pattern miner \cite{Parameter-Free-Probabilistic-API-Mining, wang2013mining, xie2006mapo} to determine the best API sequences from the returned code snippets. Following previous literature, we used Lucene \footnote{https://lucene.apache.org/} to perform a code search using a natural language query, while UP-Miner is used to analyze API usage patterns.
A text retrieval engine such as Lucene treats source code as plain text documents, which is how it builds a source code index and performs text retrieval. Pattern mining tool UP-Miner \cite{wang2013mining} analyzes snippets of codes based on the results of a Lucene search and finds API sequence patterns . In this procedure, API sequences extracted from code snippets are clustered and then frequent patterns are identified. Lastly, it clusters the frequently occurring patterns to reduce redundancy. For this experiment, we used the same code corpus as those used for evaluating PTMs and compared BLEU scores to those from PTMs.

\subsubsection{SWIM}
In \cite{swim}, Raghothaman et al. transformed natural language queries into a list of possible APIs using statistical word alignment models \cite{word-alignment-model}. Based on the predefined API list, SWIM retrieves relevant APIs. Its statistical word alignment model, however, is based on a bag-of-words assumption that does not consider how words are arranged in natural language queries and APIs are positioned in the result. Also, the model does not take contextual information into account when producing API sequences from natural language input, and therefore cannot recognize deep semantics of natural language queries. As an example, the model does not distinguish between the query 'convert txt to CSV' and 'convert CSV to txt.'

SWIM uses the statistical word alignment model to extract a list of relevant APIs based on the keywords in the given natural language query \cite{word-alignment-model} and then leverages Lucene to search for related API sequences from the extracted API list. During our experiment, we used the same dataset for evaluating SWIM as we did for evaluating PTMs.

\subsection{Deep Learning Approaches without Pretrained Knowledge}
    DeepAPI \cite{deep-api-learning} generates API usage sequences relevant to a natural language query using deep neural networks. It converts the machine translation problem into the API learning problem. DeepAPI's understanding of natural language queries is demonstrated on two fronts. Firstly, it recognizes semantically related words by embedding them in a vector representation of context instead of matching keywords. Secondly, DeepAPI learns a language query's sequence of words and the associated APIs' sequence separately instead of aligning them to each other. Therefore, DeepAPI can distinguish semantic differences between queries and generate API sequences accordingly.
    DeepAPI adapts RNN Encoder-Decoder. Using DeepAPI, a language model is trained that encodes each sequence of words (annotation) into a fixed-length context vector, and the context vector is used to decode API sequences. The model is then used to generate API sequences in response to API-related user queries.

\subsection{Deep Learning Approaches with Pretrained Knowledge}
    A more recently established and widely accepted practice in Natural Language Processing (NLP) is using Pre-Trained Language Models (PTM) and then transfer its learned knowledge to various downstream NLP tasks, such as sentiment analysis, question answering, or classification \cite{qiu2020pre}. 
    In NLP, PTMs (such as BERT) are large language models that are trained on large natural language corpora using a deep neural network in an unsupervised manner \cite{how-far-ptm}. 
    These models are then fine tuned for various downstream tasks using limited labeled datasets. 
    For example, BERT \cite{bert} is a PTM that is frequently being used for question answering and sentiment classification tasks.
    As PTMs are trained on large general domain corpora, they learn contextual linguistic information and eliminate the need to train downstream task models from scratch \cite{wada2020medicalBERT}. PTMs reduce the amount of effort (i.e., new model development time per task) to build models for each task separately, and they reduce the amount of required labeled dataset \cite{roberta}.
    Consequently, PTMs are used to transfer the learned knowledge to a new domain or a new task, and in settings where a model has not seen any example during training (known as zero-shot learning) \cite{zero}. 
    Although PTMs are used extensively and led to many advances in NLP, their applicability for API Learning and Sequence Generation is barely known \cite{EvaluatingPTM} but to what extent PTMs can be applied for API Learning problem is largely unknown.

\subsubsection{BERT}
    Devlin et al. \cite{bert} designed Bidirectional Encoder Representations from Transformers (BERT) to learn contextual word representations from unlabeled texts. Contextual word embeddings designate a word's representation based on its context by capturing applications of words across different contexts.
    BERT employed a bidirectional encoder to learn the words' contextual representations by optimizing for Masked Language Model (MLM) and Next Sentence Prediction (NSP) tasks.
    For MLM, 15\% of all the tokens are replaced with a masked token (i.e., [MASK]) beforehand, and the model is trained to predict the masked words, based on the context provided by the non-masked words.
    For NSP, the model takes sentence-pairs as input for learning to predict whether a pair-match is correct or wrong. During training, 50\% of the inputs are true consequent pairs, while the other 50\% are randomized non-consequent sentence-pairs.
    Devlin et al. trained two versions: small-sized BERT\textsubscript{BASE} and big-sized BERT\textsubscript{LARGE}. BERT\textsubscript{BASE} is a smaller model with 12 layers and 110 million parameters. BERT\textsubscript{LARGE} has 24 layers and 340 million parameters.
    BERT\textsubscript{LARGE} is more computationally expensive and consumes more memory compared to BERT\textsubscript{BASE}.
    Please note that based on the results reported in the BERT paper, BERT\textsubscript{LARGE} always exceeds BERT\textsubscript{BASE}.
    Although we have not used BERT in our study, this is the building block for the pre-trained models that we are going to discuss in this section. 
    
\subsubsection{RoBERTa} 
    Robustly optimized BERT approach (RoBERTa) outperformed all the state-of-the-art benchmarks upon release \cite{roberta}. Liu et al. modified BERT's pretraining steps that yield substantially better performance on all the classification tasks. 
    RoBERTa increased the amount of mini-batch sizes, data, and training time to train the model. 
    RoBERTa is also trained on dataset that includes longer sequences than before. The masking pattern in RoBERTa was also modified to be generated spontaneously.
    
    \textit{Reason behind choosing RoBERTa:} RoBERTa outperforms BERT on nine different NLP tasks on the GLUE benchmark. Based on these results, RoBERTa can present a reasonable choice for PTM in our study.
    
\subsubsection{CodeBERT} \label{subsec:codebert}
    CodeBERT \cite{feng2020codebert} was developed using a multilayered attention-based Transformer model, BERT \cite{bert}. As a result of its effectiveness in learning contextual representation from massive unlabeled text with self-supervised objectives, the BERT model has been adopted widely to develop large pre-trained models.
    Thanks to the multilayer Transformer~\cite{vaswani2017attention},
    CodeBERT developers adopted two different approaches than BERT to learn semantic connections between Natural Language (NL) - Programming Language (PL) more effectively.
    
    Firstly, The CodeBERT developers make use of both bimodal instances of NL-PL pairs (i.e., code snippets and function-level comments or documentations) and a large amount of available unimodal codes \cite{facos}. In addition, the developers have pre-trained CodeBERT using a hybrid objective function, which includes masked language modeling~\cite{bert} and replaced token detection~\cite{clark2020electra}. The incorporation of unimodal codes helps the replaced token detection task, which in turn produces better natural language by detecting plausible alternatives sampled from generators.
    
    Developers trained CodeBERT from Github code repositories in 6 programming languages, where only one pre-trained model is learned for all six programming languages with no explicit indicators used to mark an instance to the one out of six input programming languages. 
    CodeBERT was evaluated on two downstream tasks: natural language code search and code documentation generation. The study found that fine-tuning the parameters of CodeBERT obtained state-of-the-art results on both tasks.
    
\subsubsection{PLBART} \label{subsec:plbart}
    PLBART (Program and Language BART) \cite{plbart} is a sequence-to-sequence model capable of performing a broad spectrum of Code summarization, generation, and translation tasks. We have chosen to use PLBART in our study as it is pre-trained on a vast collection of Java and Python functions and associated NL text.
    
    The motivation behind PLBART was to develop a general-purpose model that can be used in various Program and Language Understanding and Generation (PLUG) applications. A recent development in deep learning coupled with the accessibility of large-scale Programming Language (PL) and associated Natural Language (NL) data led to the automation of PLUG applications. A fundamental feature of PLUG applications is that they require a thorough understanding of program syntax the interdependencies between the semantics of PL and NL. 
    A few research efforts in learning general purposes PL-NL representation encoders, such as CodeBERT, but that too is pretrained on a small-scale bimodal data (code-text pairs). Despite the effectiveness of existing methods, they do not have a pretrained decoder for language generation. Therefore, they still require a large amount of parallel data to train the decoder. The unified PLBART includes PL and NL pre-training on unlabeled data to learn multilingual representations applicable to many PLUG applications. 
    
    Reason for including PLBART: PLBART outperforms state-of-the-art models for code summarization, code generation, and code translation in seven programming languages. Furthermore, PLBART's effectiveness in program understanding is demonstrated by experiments on program repair, clone detection, and vulnerable code detection. 

\section{Methodology}
\label{sec:methodology}
In this section, we will discuss our dataset, experimental setup, and 
evaluation metric to \yc{measure the performance of the PTMs against the prior approaches.}

\subsection{Dataset}
We 
leverage a large scale API sequence to annotation corpus provided by Gu \yc{et.} al. \cite{deep-api-learning}, who propos DeepAPI. The description of the dataset and the collection process is described as the following.

Gu \yc{et.} al. curate a large-scale database that contains pairs of API sequences and natural language annotations. They collect 442,928 Java projects' last snapshots from GitHub. \yc{Each of the project} 
\yc{has} at least one star. After the collection, \yc{pairs of} <API sequence, annotation> are extracted by parsing source code files into Abstract Syntax Trees (ASTs) using Eclipse's JDT compiler \footnote{http://www.eclipse.org/jdt/}. 
The dependencies of an entire project repository is analyzed. All field declarations and type bindings for all classes are then documented, and all object types are replaced with the corresponding 
class types. Subsequently, the API sequence from each method body is extracted using an AST traversal.

The authors appended "APIName.new" to the API sequence after each constructor call, and "APIName.methodName" to the API sequence after each method call of a JDK class instance. When a method is called with a parameter, the parameter methods are appended before the calling method. They derive the API sequence of a sequence of statements by extracting and concatenating the API sequence of each statement consecutively. For conditional and loop statements, the authors sequentially put all the possible branches to extract APIs from the branch statements, respectively. 
Based on the Javadoc guidance \footnote{http://www.oracle.com/technetwork/articles/java/
index-137868.html}, the first sentence of a documentation comment for a method is used as a summary of the corresponding method. The authors traverse the method's AST and used the Eclipse JDT compiler to extract the method summary. 
The method summary is then used as the natural language description for the corresponding API sequence.
Following the process, the Gu \yc{et.} al. obtain a database consisting of 7,519,907 <API sequence, annotation> pairs. \yc{Gu et. al.} do not separate the dataset into training, validation, and testing sets specifically. So, we randomly selected 10,000 <API sequence, annotation> out of $\sim$500,000 pairs for the testing set and 7M for training and validation sets.

\subsection{\textbf{Experimental Setup}}
\label{experimental-setup}

\begin{figure*}[hbt!]
    \centerline{\includegraphics[width=\textwidth]{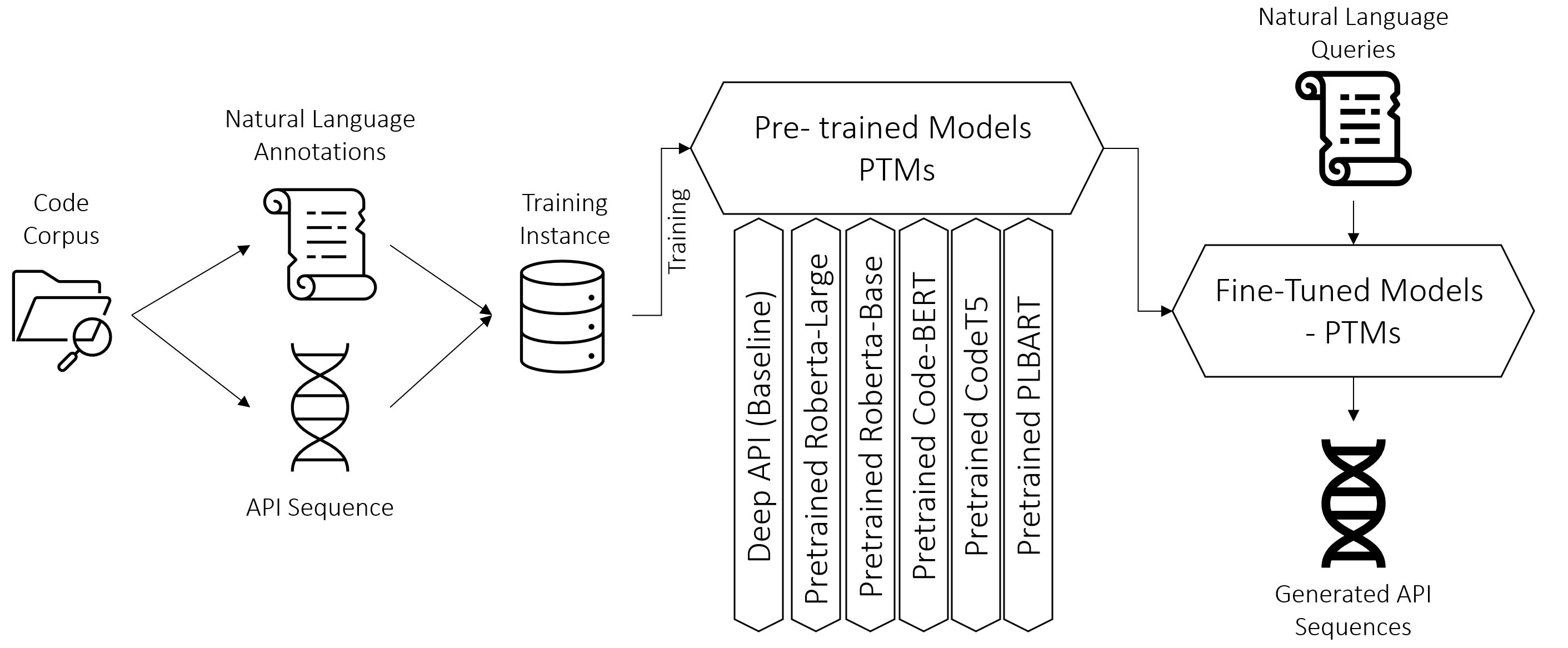}}
    \caption{Workflow of our Experiments}
    \label{fig:exp_approach}
\end{figure*}

\subsubsection{Baseline Selection:}

Following the literature, we \yc{select} three prior 
approaches 
as the baseline for evaluating \yc{the existing} PTMs. These approaches are the 
\yc{Lucene+UPMiner}, SWIM, and DeepAPI. 
\yc{We reimplement these three approaches to make sure that all experiments are performed in the same environment and leverage the same training, validation, and testing set. We believe that our implementations are correct as the results are slightly different with the results reported in the prior study \cite{deep-api-learning}.} We conjecture that the difference in performance may have resulted from using training and testing data sets that are different from the original implementation.


For Lucene+UPMiner, 
we perform the following pre-processing steps. First, we treat the Java codes that are retrieved from GitHub as plain texts. Subsequently, we remove the new line (i.e., ‘\textbackslash n’) and tab (i.e., ‘\textbackslash t’). Then, we use Lucene to index the pre-processed code snippets, and apply UPMiner to mine API usage sequences from pre-preocessed code snippets to find the frequent API usage patterns. Following \cite{deep-api-learning}, we do not fine-tune the Lucene+UP-Miner approach.
For SWIM, we train the word alignment model, and subsequently build the API indexes using the training set. Given a natural language query, SWIM leverage the word-alignment model to return a list of relevant API sequences, which is used for code synthesis in the later step. However, we only adapt the word alignment model and discard the code synthesis part as we are only interested in 
\yc{the API sequence generation task.} For DeepAPI \cite{deep-api-learning}, we leverage the original implementation provided by the authors. To train DeepAPI, we set the batch size to 200 and limit the source and target vocabularies to the top 10,000 words that are appear most frequently in the queries and API sequences.



We execute all the experiments on a Linux machine with Intel 2.21 GHz CPU and 16GB memory with two NVIDIA Tesla V100 32GB GPU.

\subsubsection{Fine-tuning PTMs:}
With the PTMs, the API sequence generation task is framed as an NL-to-PL translation problem. The PTM's encoder converts a natural language query into a context vector, and the respective PTM's decoder leverage the context vector to generate the corresponding API sequences. 

We fine-tune 
the off-the-shelf available PTMs by freezing two-thirds of the lower-level layers of the PTMs and unfreezing the rest of the upper-level layers. 
For example, for RoBERTa-Large, we freeze the 16 lower-level layers out of 24 layers following a similar setting as [2]. All the PTMs are fine-tuned with the same hyperparameter settings, e.g., the batch\_size, learning\_rate, and num\_of\_epochs are the same for all the PTMs.

As RoBERTa has been used as the codebase of most of the considered PTMs, we reuse the RoBERTa model's parameters as the starting point. 
\yc{Before feeding the input queries into the model, we also leverage the corresponding PTM's tokenizer to preprocess the input into the format that the model can understand.} The PTMs are then fine-tuned using the preprocessed data and tested on the held-out test set. We adopt the following values of hyper-parameter from the literature \cite{how-far-ptm} to fine-tune the PTMs; batch size: 16, and learning rate (AdamW): 2e-5. Further, we train the PTMs and until 50th epochs. Table \ref{table:ptm_config} shows the details and configuration of each PTM.

\begin{table}[htbp]
\caption{PTM Description}

\begin{center}
\renewcommand{\arraystretch}{1}
\resizebox{1.0\linewidth}{!}{%
\begin{tabular}{l | c | c | c | c | c}
\hline
    \textbf{Architecture} & \textbf{Used Model} & \textbf{Parameters} & \textbf{Layers} & \textbf{Hidden} & \textbf{Heads} \\
\hline
    RoBERTa & roberta-base & 125M & 12 & 768 & 12 \\
    RoBERTa & roberta-large & 355M & 24 & 1024 & 16 \\
    CodeBERT & roberta-base & 125M & 12 & 768 & 12 \\
    PLBART & bart-base & 140M & 6 & 768 & 12 \\
    Code-T5 & t5-base & 220M & 12 & 768 & 12 \\
\hline
\end{tabular}
}
\end{center}
\label{table:ptm_config}
\end{table}


\subsection{\textbf{Evaluation Metrics}}
\label{evaluation-metrics}

We use the following metric to evaluate the performance of PTMs against the prior approaches.

\subsubsection{BLEU:} We use BLEU score to evaluate how close are the generated API sequences with the ground truth. BLEU score measures the similarity of a candidate sequence to the reference sequence based on the number of n-gram hits. Equation~\ref{eq:bleu} is used to compute the BLEU score. 

\begin{equation}
\label{eq:bleu}
    BLEU = BP \cdot exp( \sum_{n=1}^{N} w_n log p_n)
\end{equation}
\begin{equation}
\label{eq:pn}
    p_n = \frac{\mbox{\# of n-grams appear in the reference + 1 }}{\mbox{\# of n-grams of candidate + 1}} \mbox{ for n = 1, ... , N}
\end{equation}
\begin{equation}
\label{eq:bp}
BP =
    \begin{cases}
      1 & \text{if c $>$ r}\\
      e^{(1-r/c)} & \text{if c $\leq$ r}
    \end{cases} 
\end{equation}
In Equation \ref{eq:bleu}, $p_n$ refers 
the n-gram hits between the candidate sequence and the reference, and $w_n$ 
equals to $1/N$
where $N$ is the number of grams to be considered. We set $N$ equals to 4. Further, $BP$ is the penalty for short candidate that may yield higher n-gram precision. $BP$ is calculated using Equation \ref{eq:bp}. In the Equation \ref{eq:bp}, $r$ and $c$ refer to the length of reference and candidate sequences.

In this paper, we express BLEU score in the percentage. A higher BLEU score indicates 
that the candidate sequence is very similar with the reference sequence. When the candidate sequence matches exactly the reference sequence, the BLEU score becomes 100\%.

\section{Results}
\label{sec:res}
We aim to answer the following research questions:
\begin{itemize}
  \item[\textbf{RQ1:}] Can PTMs yield better performance than the prior approaches?
  \item[\textbf{RQ2:}] Can different tokenization techniques \yc{boost the performance of the} PTMs?
\end{itemize}

\subsection{RQ1: Can PTMs yield better performance than the prior approaches?}

\yc{We compare PTMs with the three aforementioned approaches as the baseline. The first is Lucene+UP-Miner; a combination of Code Search and Pattern Mining where Lucene is used to perform a code search using a natural language query, and UP-Miner \cite{wang2013mining} is used to analyze the API usage patterns. The second is SWIM that leverages the statistical word alignment model to learn the API usage patterns and then uses Lucene to search for the relevant API sequences given a natural language query. The last one is DeepAPI \cite{deep-api-learning}; \yc{a deep learning-based approach that leverages} \yc{RNN-based} encoder-decoder architecture \yc{that} models} 
\yc{the API sequence generation task} as a machine-translation problem, where natural language queries are the \yc{input} and the API sequences are the \yc{output.}

\yc{Table \ref{table:baselines} shows the results of our experiments.} In Table \ref{table:baselines}, Deep-API-Auth is the author's reported BLEU score in \cite{deep-api-learning}, while Deep-API-Rep is 
\yc{the score from our experiment that leverages the authors' replication package. }
\yc{Similarly,} SWIM and Lucene+UP-Miner are also 
the scores \yc{that are obtained from our experiment.}
Table \ref{table:baselines} indicates that our results \yc{are} slightly different from the originally reported scores in \cite{deep-api-learning}. \yc{Moreover, our results also demonstrate the same trend as the prior study \cite{deep-api-learning}; DeepAPI outperforms Lucene+UP-Miner and SWIM by a significant margin.}

\begin{table}[htbp]
    \caption{\yc{Results on the} Baselines}
    
    \begin{center}
    \renewcommand{\arraystretch}{1}
    \begin{tabular}{l | c}
    \hline
        \textbf{Name of the Model} & \textbf{BLEU Score} \\
    \hline
        Deep API - Rep	&   52.25 \\
        Deep API - Auth	&   54.42 \\
        SWIM    &   17.16 \\
        Lucene+Up-Miner &   21.73 \\
    \hline
    \end{tabular}
    \end{center}
    \label{table:baselines}
    \end{table}
    
We compare the BLEU score achieved by the PTMs against the BLEU score produced by Deep-API-Rep in Table \ref{table:ptm_res}. 
\yc{Overall, the results indicates that the PTMs can outperforms Deep-API-Rep. Specifically, CodeBERT, RoBERTa-LARGE, CodeT5, and PLBART perform better than Deep-API-Rep by 3.72\%, 5.26\%, 5.82\%, and 6.93\% in terms of BLEU Score.}
    
\begin{table}[htbp]
    \caption{PTM's performance}
    
    \begin{center}
    \renewcommand{\arraystretch}{1}
\resizebox{1.0\linewidth}{!}{%
    \begin{tabular}{l | c | c | c}
    \hline
        \textbf{Name of the Model}  &   \textbf{Checkpoints}    &   \textbf{BLEU Score} & \textbf{Improvement}\\
    \hline
        RoBERTa-Large   &   50/50th epoch   &   57.88 & 5.63\%\\
        RoBERTa-Base    &   50/50th epoch   &   51.76 & -0.49\%\\
        CodeBERT    &   50/50th epoch   &   56.34 & 4.09\% \\
        Deep API - Rep &   50/50th epoch   &   52.25 \\
        \textbf{PLBART}  &   50/50th epoch   &   \textbf{59.55} & \textbf{7.3\%}\\
        Code-T5 &   50/50th epoch   &   58.44 & 6.19 \%\\
    \hline
    \end{tabular}
    }
    \end{center}
    \label{table:ptm_res}
    \end{table}
    
\vspace{3mm}
\resizebox{0.9\linewidth}{!}{%
\begin{tcolorbox}[colback=black!5!white,colframe=white!50!black,title=Findings of RQ1]
    Most 
    \yc{off-the-shelf} PTMs can outperform Deep-API-Rep by considerable margin, \yc{ranging from 3.72\% to 6.93\% in terms of BLEU score.} The \yc{PTM that is} pre-trained on natural language corpora (i.e., RoBERTa) requires bigger architecture (\yc{e.g.,} more parameters, layers, heads) to outperform Deep-API-Rep.
\end{tcolorbox}
}

\subsection{RQ2: Can different tokenization techniques boost the performance of the PTMs?}

\begin{figure}[hbt!]
    \centerline{\includegraphics[width=0.5\linewidth]{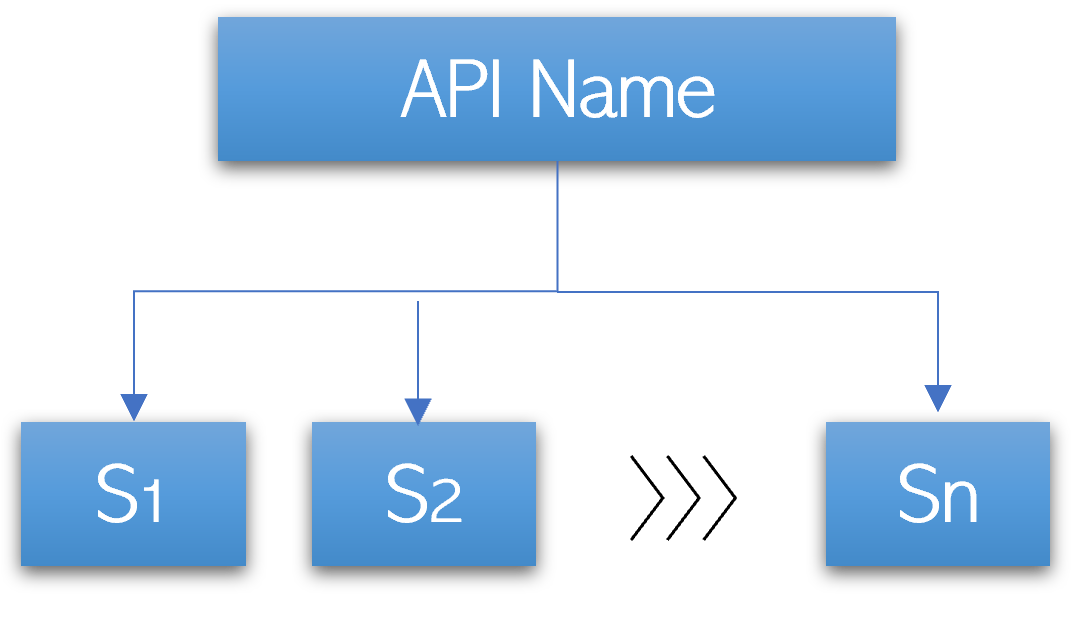}}
    \caption{Original Tokenization Approach}
    \label{fig:version_1}
\end{figure}

We implement two tokenization techniques to investigate if \yc{they can possibly help the PTMs perform better.} 
\yc{Given an API name, the default PTMs' tokenizers break down the API names into a sequence of subword tokens as illustrated in Figure~\ref{fig:version_1}.}

    In the original tokenization approach as shown in Figure \ref{fig:version_1}, the decoder tries to learn how to rearrange the tokens $S_1$, $S_2$, …, $S_n$ to 
    generate valid API names during the fine-tuning. This helps in Natural Language Processing (NLP), to make sense of the grammar and different uses of lemma in different contexts.
    But API names are constant, and we want to put more focus on the sequence, rather than API-names. Also, it is difficult for the decoder to ignore all the information these sliced token already acquired during the respective PTM’s pretraining.
    
\begin{figure}[hbt!]
    \centerline{\includegraphics[width=0.65\linewidth]{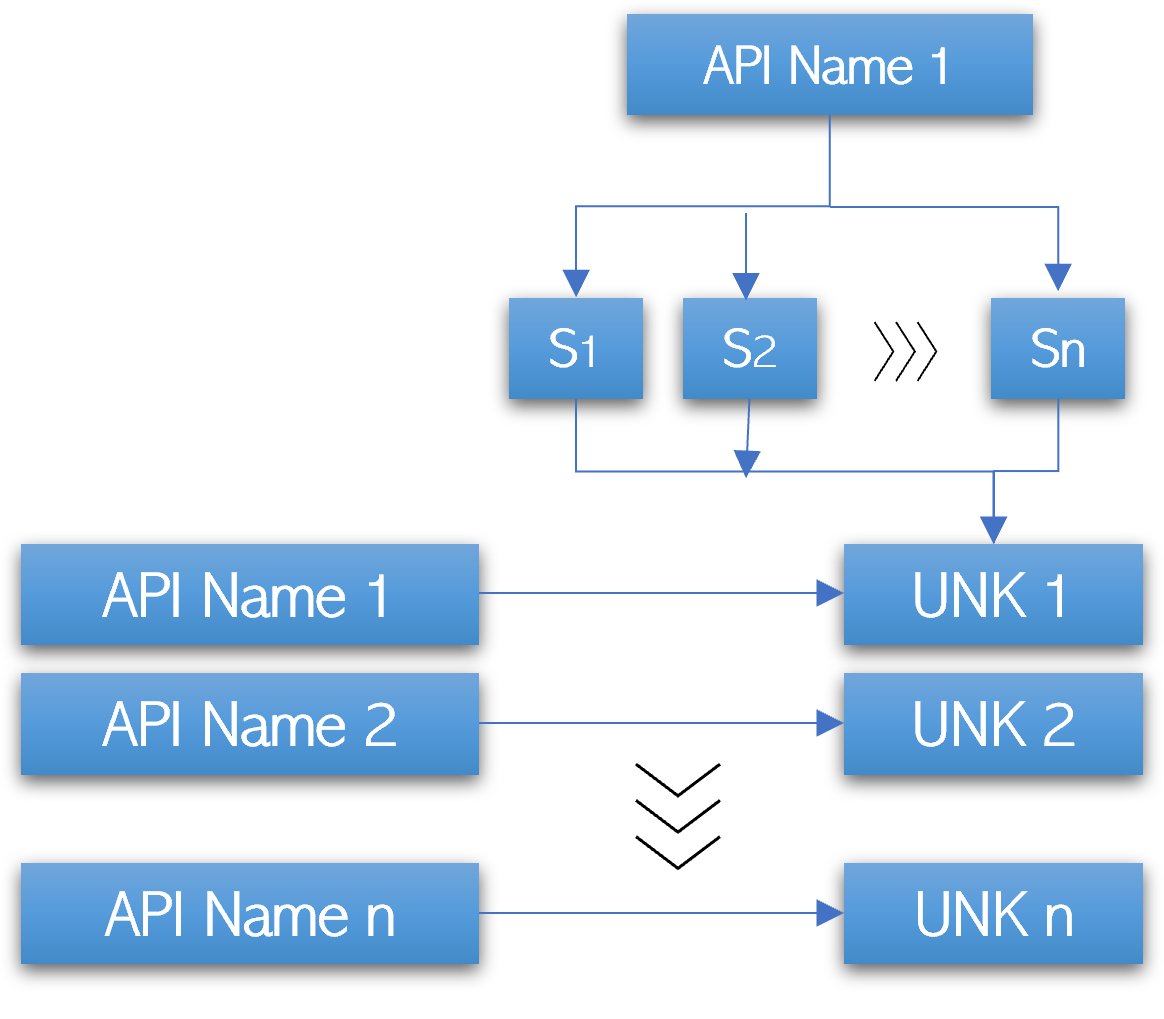}}
    \caption{Modified Tokenization Approach 1 (TA1)}
    \label{fig:version_2}
\end{figure}

    We decided to try a modified tokenization approach, where the API-names will be directly injected to vocabulary by averaging their corresponding token-slices' embeddings. Here, we let the tokenizer break down API-Names into slices; we keep track the broken down token-slices. Later, we add all the API-Names as separate Unknown embeddings to the Vocabulary, such as $<UNK-1>$, $<UNK-2>$, $\dots$, $<UNK-n>$. The Unknown tokens get initialized with the average of the embedding values of tracked token-slices. An illustration of the process is provided in Fig \ref{fig:version_2}.

From Table \ref{table:ptm_tok_1}, we observe that the Tokenization Approach 1 (TA1) \yc{enhances} the performance of the PTMs by a large margin. 
\yc{Leveraging the default tokenization}, RoBERTa-BASE could not outperform Deep-API-Rep as shown previously in Table~\ref{table:ptm_res}. After leveraging our Tokenization Approach TA1, RoBERTa-BASE along with all the other PTMs \yc{can yield better performance than} Deep-API-Rep. \yc{Specifically, }RoBERTa-BASE, CodeBERT, RoBERTa-LARGE, CodeT5, and PLBART yield 2.51\%, 8.02\%, 8.58\%, 9.23\%, and 10.46\% better BLEU Score than Deep-API-Rep respectively. \yc{Moreover, the performance of RoBERTa-LARGE, RoBERTa-BASE, CodeBERT, PLBART, and CodeT5 when using Tokenization Approach TA1 (see Table~\ref{table:ptm_tok_1}) are better by 3.32\%, 3.37\%, 4.3\%, 3.53\%, and 3.41\% in terms of BLEU score when compared against the original tokenization approach (see Table~\ref{table:ptm_res}).}

\begin{table}[htbp]
    \caption{PTM's performance: Tokenization Approach 1 (TA1)}
    
    \begin{center}
    \renewcommand{\arraystretch}{1}
    \begin{tabular}{l | c | c}
    \hline
        \textbf{Name of the Model}  &   \textbf{Checkpoints}    &   \textbf{BLEU Score} \\
    \hline
    RoBERTa-Large   &   50/50th epoch   &   61.2	\\
    RoBERTa-Base    &   50/50th epoch   &   55.13	\\
    CodeBERT    &   50/50th epoch   &   60.64	\\
    Deep API - Rep &   50/30th epoch   &   52.25	\\
    \textbf{PLBART}  &   50/50th epoch   &   \textbf{63.08}	\\
    Code-T5 &   50/50th epoch   &   61.85	\\
    
    \hline
    \end{tabular}
    \end{center}
    \label{table:ptm_tok_1}
    \end{table}
    
    The Tokenization Approach 1 (TA1) 
    can introduce a problem as the 
    \yc{subword tokens} retain a lot of natural language information from the pre-training corpora. 
    Such an unrelated information 
    is later contributes to 
    \yc{the embedding representation of the whole API names}, which we conjecture to have induced some noises towards
    the representation.
    
\begin{figure}[hbt!]
        \centerline{\includegraphics[width=0.5\linewidth]{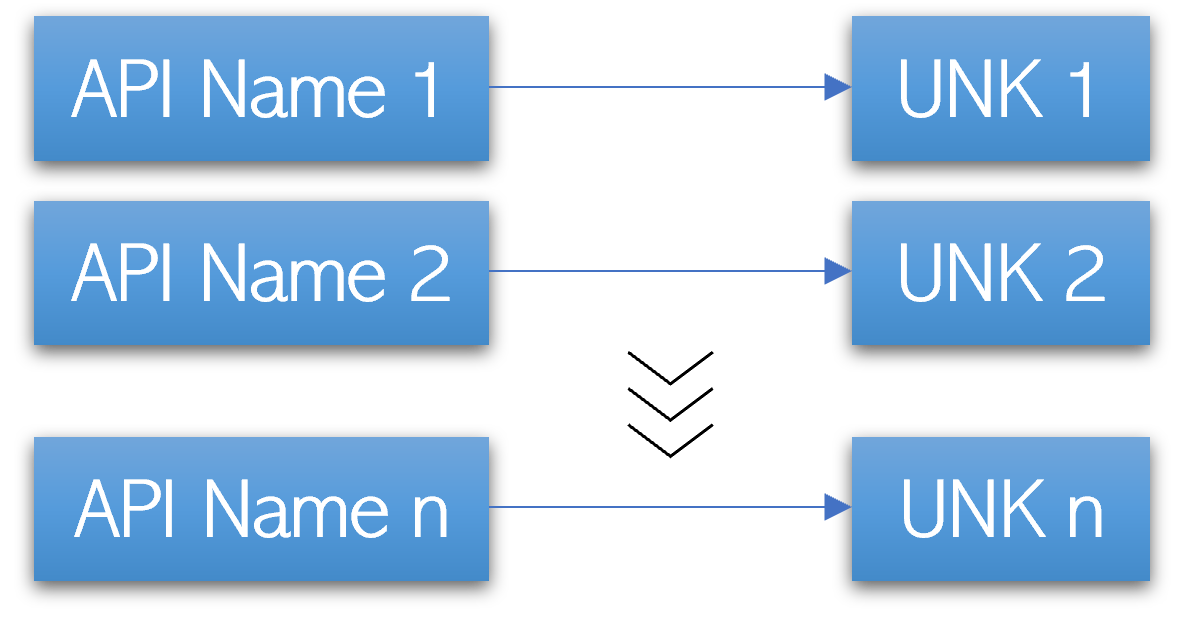}}
        \caption{Modified Tokenization Approach 2 (TA2)}
        \label{fig:version_3}
    \end{figure}

    To circumvent the problem, we propose the second Tokenization Approach TA2, where 
    \yc{the tokenizer does not break down the API names into }
    \yc{subword tokens}. 
    First, we add all the API-Names (e.g., $<API-1>$, $<API-2>$, $\dots$ shown in Figure~\ref{fig:version_3}) as separate 
    tokens to the vocabulary. $<API-n>$. 
    \yc{These API names are then converted the corresponding vector representations (<UNK{1,2,…,n}> in Figure~\ref{fig:version_3}). These representations are one-hot encoded vectors.}
    \yc{Intuitively, Tokenization Approach 2 (TA2) can help the decoder to put more focus on the API sequences instead of learning to rearrange the subword tokens.}
    
\begin{table}[htbp]
    \caption{PTM's performance: Tokenization Approach 2 (TA2)}
    
    \begin{center}
    \renewcommand{\arraystretch}{1}
    \begin{tabular}{l | c | c}
    \hline
        \textbf{Name of the Model}  &   \textbf{Checkpoints}    &   \textbf{BLEU Score} \\
    \hline
        RoBERTa-Large   &   50/50th epoch   &   62.25 \\
        RoBERTa-Base    &   50/50th epoch   &   55.81 \\
        CodeBERT    &   50/50th epoch   &   61.51 \\
        Deep API - Repl &   50/50th epoch   &   52.25 \\
        PLBART  &   50/50th epoch   &   63.19 \\
        Code-T5 &   50/50th epoch   &   62.31 \\
    
    \hline
    \end{tabular}
    \end{center}
    \label{table:ptm_tok_2}
    \end{table}
    
    From Table \ref{table:ptm_tok_2}, the results indicates that Tokenization Approach 2 (TA2) \yc{can yield slightly better performance than when using Tokenization Approach - 1 (TA1)}. \yc{Specifically, }RoBERTa-BASE, CodeBERT, RoBERTa-LARGE, CodeT5, and PLBART produced 1.05\%, 0.68\%, 0.87\%, 0.11\%, and 0.46\% better BLEU Score than their implementation with TA1, respectively. 
    
\vspace{3mm}
\resizebox{0.9\linewidth}{!}{%
\begin{tcolorbox}[colback=black!5!white,colframe=white!50!black,title=Findings of RQ2]
    \yc{Our} Tokenization Approaches can help PTMs to better learn API-Sequences. \yc{Both Tokenization Approach 1 (TA1) and Tokenization Approach 2 (TA2) perform better than the default tokenization, as shown in Table~\ref{table:ptm_tok_1} and Table~\ref{table:ptm_tok_2}.} Preventing the tokenizer from decomposing API names into 
    subword tokens can help improve PTMs' performance, if handled properly.
\end{tcolorbox}
}
\section{Discussion}
\label{sec:disc}

This section details the lessons we learned from our experiments and discusses threats to their validity.

\subsection{Lessons Learned} 
\vspace{2mm}
\noindent\textbf{Reasons behind PTMs generally performing better}

The semantic gap between API 
\yc{sequences} and natural language descriptions presents a major challenge for API learning. Generally, information retrieval approaches assume that the natural language and code are made up of a bag of words without adequately understanding the high-level semantics of the language. In contrast, the existing deep learning-based approaches focus on two distinct things during pre-training: \yc{1)} learning the semantics of natural language and programming language, and \yc{2)} understanding how the model can 
\yc{correlate between the natural language queries and the intended API sequences.} However, the PTMs have already accumulated the semantic knowledge of natural and programming languages \yc{in the pre-training phase}. Consequently, when PTMs \yc{are finetuned using }the $\langle$API Sequence, Annotation$\rangle$ pairs, instead of focusing on the semantics of natural and programming languages and how given queries contribute to the 
\yc{generation} of the corresponding API sequences \yc{at the same time}, they can only focus on the latter task to be more effective.

To achieve better 
performance, the PTMs rely on three distinct advantages of deep learning-based models as well as the retained NL or NL-PL linguistic knowledge \yc{from the pre-training phase}. These advantages are: using \yc{contextualized }word embeddings to make semantic sense of the natural language queries, understanding the 
\yc{sequences} order of the queries instead of \yc{treating it as a} bag-of-words, and generating a translated sequence instead of searching specific samples for the correct arrangement of output APIs \cite{deep-api-learning}. 

\vspace{2mm}
\noindent\textbf{Fine-tuning PTMs \yc{pre-trained on NL corpora} with large architecture can help \yc{to} learn API better}

From Table \ref{table:ptm_res}, we observe that between two PTMs pre-trained on Natural Language (NL) corpora, i.e., RoBERTa-BASE and RoBERTa-LARGE, the latter outperforms the baseline \yc{Deep-API-Rep, unlike} the BASE model. So, it is recommended that researchers, developers, or practitioners should go for fine-tuning the larger architecture of any given NL pre-trained PTM for achieving better performance. The same performance as the LARGE models can also \yc{possibly} be achieved by tweaking the parameters during the pre-training of the BASE models. Researches in the neural machine translation field has shown that small mini-batches combined with the appropriate learning rate and step size \cite{ott2018scaling} can improve 
\yc{the performance of a deep learning model}. \yc{Prior study} \cite{roberta} also confirms that RoBERTa is also amenable to large batch training and proportionate step size and learning rate. If tweaking parameters during pre-training of the smaller models is not an option, we recommend using bigger architectures instead.

\vspace{2mm}
\noindent\textbf{Fine-tuning PTMs \yc{pre-trained on NL-PL corpora}, with those with small architecture, can yield better performance than those that are pre-trained only on NL corpora.}

We also found that PTMs pre-trained on NL-PL corpora (e.g., CodeBERT, CodeT5, and PLBART) can outperform the \yc{other PTMs that are pre-trained only on NL corpora.}
We conjecture that this performance boosts stem from to the bimodal (NL-PL) data and code-related objective tasks \yc{that are used to pre-train the the models in the pre-training phase.} 

\vspace{2mm}
\noindent\textbf{Different tokenization approaches can possibly boost the performance of the PTMs}

    As part of the original tokenization approach, the decoder learns to rearrange predefined tokens to create the API name during \yc{the} fine-tuning \yc{phase}. This approach helps Natural Language Processing (NLP) tasks by making sense of grammatical structure and the different uses of the lemma in different contexts; however, API names are invariant in our case, and we prefer to focus on the sequence generation rather than API names. 
    
    Our Tokenization Approach 1 (TA1) lets the tokenizer break down API names into 
    \yc{subword tokens} while tracking 
    \yc{to which API names each token belongs}. Later, we add all the API names as separate Unknown embeddings to the Vocabulary, which get initialized with the average embedding values of the tracked 
    \yc{subword tokens}. Tokenization Approach 1 (TA1) enhances the performance of the PTMs by a large margin. \yc{However, } this approach is not completely without disadvantages. In Tokenization Approach 1 (TA1), \yc{those subword tokens} retain a lot of textual information from their respective pre-training corpora, which later causes some noise in the API representations. Therefore, we propose a simpler tokenization method that initially hides API-Names from the tokenizer and then adds all API-Names to the Vocabulary as one-hot encoded embeddings. Overall, the PTMs are marginally can perform better when using this approach (TA2) than when they use Tokenization Approach (TA1).
    
\vspace{2mm}
\noindent\textbf{Pre-training and objective tasks of the PTMs play a huge role in downstream tasks}

\yc{Our study leverage a dataset containing Java API sequences that is curated from GitHub.} 
All PTMs are pre-trained on NL-PL corpora, except for RoBERTa-BASE and LARGE that are pre-trained on only NL corpora. 
\yc{Our results demonstrate that the PTM pre-trained on NL-PL corpora can perform better than the PTMs pre-trained on only NL corpora. Such a finding also aligns with the prior study~\cite{plbart}. Hence, researchers and practitioners should pay special attention before selecting PTMs that have been pre-trained on unrelated domains.} The closer the pretraining domain is for the PTMs, the better PTMs can perform for the downstream tasks. Also, not all readily available PTMs make their decoder open-sourced and not all PTMs are pre-trained on a machine translation objective task. As we frame our API Sequence generation task to a machine translation task from natural language to code, the PTMs that have already pre-trained on machine translation task can perform better.
    
\vspace{2mm}
\noindent\textbf{Modified tokenization approach help all the PTMs uniformly}

With the proposed tokenization techniques, all PTMs gain almost equal performance gain for the downstream API learning task. We can observe that all PTMs shows a general improvement in their performance and maintained their original ranking, even after implementing the modified tokenization approach. We hypothesize that their underlying common Transformer model, uniform parameter settings, and architecture are responsible for the almost consistent increase in performance.  

\subsection{Threats to Validity}

There are several threats to validity identified in this study. As we have adopted different tools, dataset, approaches, and methods from the literature, we also assume their latent threats to validity:

\vspace{2mm}
\noindent\textbf{A wide range of PTMs not evaluated}

According to Microsoft's CodexGLUE leader-board \footnote{https://microsoft.github.io/CodeXGLUE/}, three other PTMs can better serve our purpose: Text2Java-BigJavaDataSet, CoText, Text2Java-T5. But due to the lack of open-sourced implementation, we could not incorporate Text2Java-BigJavaDataSet and Text2Java-T5. We have contacted the authors of CoText \cite{cotext}, but we are yet to receive a response.

\vspace{2mm}
\noindent\textbf{Narrow API domain}

Our paper only incorporated one dataset from a previous study on Deep-API Learning, which only examined APIs and related projects from the JDK API. Consequently, the dataset may lack generalizability and fair representation of other libraries and programming languages. But it can very well work as a reference to show the potential of a designed approach for guiding the researchers and practitioners.

\vspace{2mm}
\noindent\textbf{Annotation quality}

Annotations of API sequences are collected from the first sentence of comments in the documentation, which may contain noise. The dataset curators did not consider the fact that some other sentences may also provide useful information. In the future, we plan to curate a more comprehensive dataset that will include more diverse programming languages and more informative annotations. 

\vspace{2mm}
\noindent\textbf{Size of the PTMs}

How large should the PTM be to start seeing improvements over non-PTM approaches is not studied in our work but this is a great research question and we plan to investigate this for our future work. We need to vary the size of the PTM and investigate the performance for each size change. As the experiment will require a substantial amount of time, we plan to integrate this in our future work.

\vspace{2mm}
\noindent\textbf{Data Leakage}

We have not checked if the \textit{<Natural Language Annotation, API Sequence>} test set were ever a part of the considered PTMs’ pretraining. We only ensured that no test data leaked into the training dataset that we used to fine-tune the PTMs by cross-checking the \textit{<Natural Language Annotation, API Sequence>} pairs.
The involved PTMs have utilized different code sources collected over different timelines. Although all the papers that proposed the PTMs explicitly mention the sources, these repositories were not made readily available. As we need to track, collect, and process these repositories by going through the research papers and related documentation, this task will take a lot of time to implement and cross-check. Moreover, even if some codes or natural language comments associated with our training dataset were present during the pretraining of these PTMs, these codes and comments were used to achieve a different objective task that focused on learning something different from API learning.

\vspace{2mm}
\noindent\textbf{Imbalanced Training and Testing Data}

    A possible internal validity can be related to the training and testing split in the dataset. We randomly selected 10,000 <API sequence, annotation> out of $\sim$500,000 pairs for testing and 7M for training and validation data. Due to time and resource constraints, we did not use stratified k-fold cross-validation to avoid the bias that might be introduced to the results by the test set. 
    In addition, to mitigate threats to the validity of the results, we kept hyper-parameter values the same for all PTMs in all the steps of experiments. We ran all the experiments on a single machine, and we reported the machine configuration to enforce the reproducibility of the results. Furthermore, we consider the same metrics to compare the PTMs with prior approaches. 

\vspace{2mm}
\noindent\textbf{Evaluation metric}

We relied on a single evaluation metric, BLEU Score. Although this evaluation metric has been heavily used in literature to evaluate machine translation, we could incorporate an evaluation metric based on manual human validation of the translation tasks. In the future, we will report different variations of BLEU scores and human evaluation of API Sequence generation task.

\vspace{2mm}
\noindent\textbf{Construct Validity}

    The selection of the prior approaches and PTMs can pose a validity threat to our study. We identified the three most common approaches as priors by examining the highly practiced methods, tools, and techniques employed by researchers and application developers. We chose these methods as PRIORs after conducting a comprehensive literature review.
    The PTMs were adopted by following previous studies that performed Code to Text conversion-related tasks with PTMs. 

\section{Related Work}
\label{sec:related_work}
Our work is closely related with code generation and code retrieval tasks. We describe some related work as follows.

\vspace{2mm}
\noindent\textbf{Deep Learning-based Code Generation.} 

Various deep learning-based approaches have been proposed to address code generation tasks. Ling et. al.~\cite{DBLP:conf/acl/LingBGHKWS16} proposed Latent Predictor Network (LPN) to translate the natural language to Java and Python languages. LPN represent the output as code tokens at the character granularity and applies copy-mechanism to deal with unknown tokens. A number of studies improve the performance of LPN by incorporating code structural information. Some approaches that belong to this category are Abstract Syntax Network (ASN)~\cite{DBLP:conf/acl/RabinovichSK17} and Syntactic Neural Model (SNM)~\cite{DBLP:conf/acl/YinN17}. Both approaches represent the output as ASTs instead of sequences of code tokens. These two approaches can perform better than LPN because it leverages the grammar of the AST to constraint the search space of the output. Patois~\cite{DBLP:conf/nips/ShinABP19} adopts Syntactic Neural Model~\cite{DBLP:conf/acl/YinN17} and leverages pattern mining technique to improve its performance. The idea is to generate a frequent pattern (a common code fragment) at one time step instead of generating it token by token in several time steps. Moreover, Sun et. al.~\cite{DBLP:conf/aaai/SunZXSMZ20} propose a transformer-based architecture specialized for code generation task called TreeGen.

\vspace{2mm}
\noindent\textbf{Deep Learning-based Code Search.} 

Some approaches have been proposed to address code search task. Gu et. al.~\cite{DBLP:conf/icse/GuZ018} proposes DeepCS, a deep-learning approach that maps natural language and code snippets into a shared vector space. The main idea of DeepCS is by training the network to map a pair of relevant natural language descriptions and its code snippet such that both are mapped close to each other in the vector space. There are various similar approaches that are proposed with the same underlying idea, such as UNIF\cite{DBLP:conf/sigsoft/CambroneroLKS019}, MMAN~\cite{DBLP:conf/icsm/GuCM21}, CDRL~\cite{DBLP:conf/wcre/HuangQZW20}, CQIL~\cite{DBLP:conf/icsm/LiQYSC20}, and CARLCS~\cite{DBLP:conf/iwpc/ShuaiX0Y0L20}. The main difference between those approaches are the code features (e.g., sequence of code tokens, method names, method signature) used to train the network and the network architecture.

\vspace{2mm}
\noindent\textbf{PTM for Sequence Generation.}

\cite{cpt} proposed a novel Chinese Pre-trained Unbalanced Transformer (CPT) that is designed for both natural language understanding (NLU) and natural language generation (NLG) tasks. It consists of three parts: a shared encoder, an understanding decoder, and a generation decoder. Two specific decoders with a shared encoder are pre-trained with masked language modeling (MLM) and denoising auto-encoding (DAE) tasks, respectively. CPT greatly accelerates the inference of text generation. 
\cite{cao2021answer} tried to help app developers generate answers that are related to the users’ issues. The app response generation models use deep neural networks and train on large amount of user feedbacks for training.
\cite{edunov2019fos} examined different strategies to integrate pre-trained representations into sequence to sequence models and apply it to neural machine translation and abstractive summarization. The study find that pre-trained representations are most effective when added to the encoder network which slows inference by only 14
\cite{mass} get inspired by the success of BERT and proposed MAsked Sequence to Sequence pre-training (MASS) for the encoder-decoder based language generation tasks. It adopts the encoder-decoder framework to reconstruct a sentence fragment given the remaining part of the sentence: its encoder takes a sentence with randomly masked fragment as input, and its decoder tries to predict this masked fragment so that it can jointly train the encoder and decoder to develop the capability of representation extraction and language modeling. 
\cite{chi2020cross} focused on transferring supervision signals of natural language generation (NLG) tasks between multiple languages. The study proposed to pretrain the encoder and the decoder of a sequence-to-sequence model under both monolingual and cross-lingual settings. The pre-training objective encourages the model to represent different languages in the shared space, so that it can conduct zero-shot cross-lingual transfer. 
\cite{bart} is a denoising autoencoder for pretraining sequence-to-sequence models, which is trained by corrupting text with an arbitrary noising function, and learning a model to reconstruct the original text. It uses a standard Tranformer-based neural machine translation architecture.
\cite{ernie-gen} proposed an enhanced multi-flow sequence to sequence pre-training and fine-tuning framework named ERNIE-GEN, which bridges the discrepancy between training and inference with an infilling generation mechanism and a noise-aware generation method. To make generation closer to human writing patterns, this framework introduces a span-by-span generation flow that trains the model to predict semantically-complete spans consecutively rather than predicting word by word. 
\cite{gan} trained GANs for language generation that has proven to be more difficult, because of the non-differentiable nature of generating text with recurrent neural networks. This study showed that recurrent neural networks can be trained to generate text with GANs from scratch using curriculum learning, by slowly teaching the model to generate sequences of increasing and variable length. 
\cite{deltaLM} introduced DeltaLM, a pretrained multilingual encoder-decoder model that regards the decoder as the task layer of off-the-shelf pretrained encoders. Specifically, it augment the pretrained multilingual encoder with a decoder and pre-train it in a self-supervised way. To take advantage of both the large-scale monolingual data and bilingual data, DeltaLM adopts the span corruption and translation span corruption as the pre-training tasks.

\vspace{2mm}
\noindent\textbf{PTM for Code Generation and Understanding}

\cite{shin2021survey} categorized approaches to generating source code from natural language descriptions by their input and output forms. The study also suggested the future direction of this research domain to improve automatic code generation using natural language by analyzing the current trend of approaches. 
\cite{poesia2022synchromesh} proposes Synchromesh, a framework to improve the coding reliability of pre-trained models. Using Target Similarity Tuning, this framework retrieves a few-shot example from a training bank. The second stage involves feeding examples to a pre-trained language model, which samples programs based on Constrained Semantic Decoding.
\cite{Li2022PL} developed AlphaCode, a system for generating code that can create novel solutions to problems that require deeper reasoning. 
\cite{perez2021codeGen} explores the possibility of applying similar techniques to a highly structured environment with strict syntax rules. Based on pre-trained language models, the study proposes an end-to-end machine learning model for code generation in Python.
\cite{Svyatkovskiy2020IntelliCode} has introduced IntelliCode Compose - a general-purpose multilingual code completion tool that can generate up to entire lines of syntactically correct code from sequences of code tokens of arbitrary types. 1.2 billion lines of Python, C\#, JavaScript, and TypeScript source code were used to train a state-of-the-art generative transformer model. 
\cite{xu2020External_Knowledge} studied the effectiveness of incorporating two forms of external knowledge: automatically mined NL-code pairs from the programming QA forums Stack Overflow and API documentation of programming languages. 
\cite{qi2021prophet} presented ProphetNet, a pre-training-based method for generating English text summaries and questions. ProphetNet was also extended into other domains and languages to create the ProphetNet family pre-training models, named ProphetNet-X, where X can be English, Chinese, Multi-lingual, and so on. Researchers have pretrained ProphetNet-Multi, ProphetNet-Zh, ProphetNet-Dialog-En, and ProphetNet-Dialog-Zh cross-lingual dialog generation models.

\section{Conclusion and Future Work}
\label{sec:concluson}

In this paper, we apply Pre-trained Transformer based Models (PTM) for constructing API usage sequences for an API-related natural language query. According to our empirical study, PTMs perform better, beat the state-of-the-art approaches, and effectively produce API sequences based on natural language queries. Although PTMs have shown promise in Natural Language and Programming Language translation tasks, we are the first to demonstrate their effectiveness in API learning. Additionally, we investigated different tokenization methods that improved the performance of PTMs.
Other software engineering problems such as code search and fault localization in API sequence may benefit from our findings in this paper. 

For future work, we will empirically study more PTMs and determine their best performance for API learning under different custom hyper-parameters tuning. We will incorporate more libraries from different programming languages to ensure a more fair and generalized dataset. Also, we can adopt a more sophisticated approach to generate annotation than just focusing on the first line of method-level comments. 
We will also explore the applications of PTMs to code search and fault localization in API sequence problems and investigate the synthesis of sample code from API sequences.

\begin{acks}
This research / project is supported by the Ministry of Education, Singapore, under its Academic Research Fund Tier 2 (Award No.: MOE2019-T2-1-193). Any opinions, findings and conclusions or recommendations expressed in this material are those of the author(s) and do not reflect the views of the Ministry of Education, Singapore.
\end{acks}

\balance
\bibliographystyle{ACM-Reference-Format}
\bibliography{sample-base}

\end{document}